\documentstyle[prd,aps,epsf]{revtex}
\begin{document} 
\oddsidemargin=0.5cm
\begin{flushleft}
 {\bf{\it  Nature, 5 September 1996}}

\bigskip

{\Large{\bf 
Self-gravity as an explanation of the fractal structure of the interstellar 
medium }}\\
\medskip
\noindent
H.J. de Vega$^{1}$, N. S\'anchez$^{2}$, F. Combes$^{2}$ \\
\medskip
\noindent
$^{1}$Laboratoire de Physique Th\'eorique et Hautes Energies, Universit\'e
Paris VI, Tour 16, 1er \'etage, 4 Pl. Jussieu 75252 Paris, France \\
$^{2}$DEMIRM, Observatoire de Paris, 61 Av. de l'Observatoire, 75014 Paris,
 France \\
\end{flushleft}
\bigskip

\noindent
{\bf 
The gas clouds of the interstellar medium have a fractal structure,
the origin of which has generally been thought to lie in turbulence
\cite{larson}-\cite{fal}.
The energy of turbulence could come from galactic rotation    
at large-scale, then cascade down to be dissipated on small-scales by 
viscosity \cite{obser,fle}; it has been suggested that such turbulence
helps to prevent massive molecular clouds from collapsing in response
to their own gravity \cite{hen,chi}.
Here we show that, on the contrary, self-gravity itself may be
the dominant factor in making clouds fractal.
We develop a field-theory approach to the structure of clouds, 
assuming them to be isothermal, and with only gravitational
interactions; we find that the observed fractal dimension of the clouds
arise naturally from this approach. Although this result does not imply
that turbulence is not important, it does demonstrate that the fractal
structure can be understood without it.
}
\medskip

The interstellar medium (ISM) is an ensemble of gas clouds and dust, composed 
mainly of hydrogen (either atomic HI, or molecular H$_2$) and 
helium (25\% by mass), with other elements present in trace 
amounts (dust is only 1\% in mass). The bulk of the ISM is 
distributed in cold clouds (T $\sim$ 5-50 K), forming a very 
fragmented and clumpy structure, confined to the galactic plane of 
spiral galaxies. 

For at least two decades, radioastronomy line observations (HI 
at 21cm wavelength, and CO 
at a wavelength of 2.6mm for the major lines), 
have told us that the ISM is composed of 
a hierarchy of structures, with masses from about 1 
M$_\odot$ to 10$^6$ M$_\odot$. Structures have been observed in the 
ISM with sizes from 10$^{-4}$pc (20 AU or 3 10$^{14}$cm) to 100pc. The largest 
of these structures are
giant molecular clouds or complexes, thought to 
be the largest self-gravitating structures in the Galaxy. Above 100pc, 
larger structures would be destroyed by galactic shear. The 
accumulation of observations at many scales, and with many tracers of 
the ISM (CO and its isotopes, HCN, CS, NH$_3$, HI, dust as in 
Fig. 1) revealed that 
the interstellar medium obeys power-law relationships between 
size (R), mass (M) and internal velocity dispersion ($\Delta$v) (see
for example \cite{larson,obser}):
\begin{equation}\label{vobser}
M (R)  \sim    R^{d_H}     \quad  ,   \quad  \Delta v \sim R^q \; ,
\end{equation}
These apply across the
entire observed range of structure sizes and masses, with 
Haussdorf dimension ($d_H$) and the power $q$
\begin{equation}\label{expos}
1.5    \leq   d_H    \leq   2. \quad , \quad \; 0.3  \leq   q  \leq  0.5 \; . 
\end{equation}
Structures appear virialised at any scale: 
the scaling laws obey the relationships 
$ \Delta$v$^2$ $\propto$ GM/R, or equivalently $q =(d_H -1)/2$.
\bigskip

Here we apply, for the first time,
field theory and Wilson's approach to critical phenomena \cite{kgw},
to the problem of a gravitational gas in statistical equilibrium.
We consider a gas of non-relativistic particles 
in thermal equilibrium at temperature $T$
interacting with each other
 through Newtonian gravity.
We  work in the grand canonical ensemble, allowing for a variable
number of particles $N$.
The grand partition function of the system of particles, of mass
$m$ and phase space coordinates $p$ and $q$, can be written as
\begin{equation}\label{gfp}
{\cal Z} = \sum_{N=0}^{\infty}\; {{z^N}\over{N!}}\; \int\ldots \int
\prod_{l=1}^N\;{{d^3p_l\, d^3q_l}\over{(h)^3}}\; e^{- {{H_N}\over{kT}}}, \quad
{\rm with} \quad 
H_N = \sum_{l=1}^N\;{{p_l^2}\over{2m}} - G \, m^2 \sum_{1\leq l < j\leq N}
{1 \over { |{\vec q}_l - {\vec q}_j|}}
\end{equation}
$G$, $h$ and $k$ are Newton, Planck and Boltzmann constants respectively
and $z$ is the fugacity $z= e^{{{\bar \mu}\over{kT}}}$;
$ {\bar \mu} $ is  the gravito-chemical potential.
  Transforming this expression through a functional integral
\cite{origen,stra}, it can be
shown that this system is exactly equivalent to the
theory of a scalar field $\phi({\vec x})$
(the detailed derivation will be found in de Vega, S\'anchez \& Combes, 1996,
in prep.); 
the grand canonical partition function
$ {\cal Z} $ can be expressed as a functional integral
\begin{eqnarray}\label{zetafi}
{\cal Z} &=&  \int\int\;  {\cal D}\phi\;  e^{-S[\phi]} \; , \quad
{\rm with} \quad 
S[\phi]  \equiv   {1\over{T_{eff}}}\;
\int d^3x \left[ \frac12(\nabla\phi)^2 \; - \mu^2 \; 
e^{\phi({\vec x})}\right] \; , \cr \cr
 T_{eff} &=& 4\pi \; {{G\; m^2}\over {kT}} \quad , \quad
\mu^2 = {{\pi^{5/2}}\over {h^3}}\; z\; G \, (2m)^{7/2} \, \sqrt{kT} \; ,
\end{eqnarray}
The parameter $\mu$ coincides
with the inverse of the Jeans length 
$\mu = \sqrt{{12}\over{\pi}} {{1}\over{d_J}}$.
The stationary points of the action $ S $ are given by the equation
for the undimensioned field $\phi$:
\begin{equation}\label{simple}
\Delta\phi = -\mu^2 e^{\phi}\; ,
\end{equation}
which is exactly the equation satisfied by the gravitational field
$ U = - { kT \over m } \phi $ of a perfect isothermal gas. Indeed, the
equation of state combined 
with the hydrostatic equilibrium equation yields 
$\rho = \rho_0 \; e^{-\frac{m}{kT} U}$. The application of the Poisson
equation leads then to (\ref{simple}), provided that 
$\rho_0 = z (2\pi m kT h^{-2})^{3/2} $.
This leads to the well-known solution of the isothermal sphere,
and small fluctuations around the stationary point have been
studied in \cite{kh}. 
In terms of the scalar field $\phi$, the particle density 
can be expressed as 
$  <\rho({\vec r})> =  -{1 \over {T_{eff}}}\;<\Delta \phi({\vec r})>=
{{\mu^2}\over{T_{eff}}} \; <e^{\phi({\vec r})}> $
where $ <\ldots > $ means functional average over  $ \phi $
with statistical weight $  e^{-S[\phi]} $.

The  term $-\mu^2 e^{\phi}$  which makes $ S $ unbounded from
below reflects the short-distance gravitational attraction.
We limit the newtonian forces at short distances 
since there the interparticle interaction is no more purely gravitational.
 The ISM in isothermal conditions (i.e. the ISM
 in contact with the heat bath represented by the cosmic background 
radiation), is unstable through Jeans instability at any scale \cite{pc}; 
fragmentation occurs down to the scale where
the coupling with the thermal bath breaks down, at which the
regime becomes adiabatic instead of isothermal. This scale is that of
the smallest possible fragments. Our isothermal gravitational gas model has
 therefore a natural cutoff here.  

Now that we have derived the scalar field representing the problem of
$N$-body in gravitational interaction, we work in two directions.
 We study first the perturbative method, then the
renormalization group approach. 

\bigskip

First we can note that $ S[\phi] $ has no
constant stationary points, except $\phi_0 = -\infty$. 
In order to compute  perturbations around $\phi_0$, we  add a small
constant term $\delta$ in the density, so that $\phi_0 =\log \delta$ 
is finite for non zero values of the constant $\delta$ (de Vega,
S\'anchez \& Combes, 1996, in prep). 
The perturbative development in terms of the dimensionless coupling
constant $ g = \sqrt {\mu T_{eff}} $ reveals that the 
the field $\phi$ is  massless;  the two-points density
correlation function decreases as $ r^{-2} $ for large distances.

Therefore, the theory remains critical, for a large
range of values of the physical parameters.
Since we consider the gas inside a large sphere of radius $ R $ 
($ R \leq 100\, pc $,  since other forces are involved above such scale) no
divergences appear at large radii.
More information is gained when 
the perturbative development is made around the spherically
symmetric solution is $\phi^c = \log {{2} \over {\mu^2 r^2} }$, which 
is invariant under scale transformations. 

\bigskip

The next step to analyze this theory is to use the renormalization group.
This non-perturbative approach is the most powerful framework to derive 
scaling behaviours in field theory (e.g. \cite{kgw,dg,ll}).
As is well known, the  correlation length $\xi$ for infinite volume systems 
becomes infinite at criticality, as $\xi \sim \Lambda^{-\nu}$
when $\Lambda \to 0$; $\Lambda ={{\mu^2}\over{T_{eff}}}$ is the distance to 
the critical point ($\Lambda = T-T_c$ in condensed matter and spin systems) 
and varies according to the renormalization group transformations.
Since  here our system is critical on a finite size $ R $, it is not
singular and we have $ \xi \sim R $, i.e. $\Lambda \sim R^{-\frac{1}{\nu}}$.
The mass density $ m \, \rho \sim e^{\phi} $ is identified  
with the energy density of the renormalization group (also called thermal
operator).
 The partition function can be written as
\begin{equation}\label{Zsca1}
{\cal Z}(\Lambda) = 
 \int\int\;  {\cal D}\phi\;  e^{ -S^* + \Lambda
\int d^3x  \; e^{\phi({\vec x})}\;}\; ,
\end{equation}
where $S^*$ stands for the action at the critical point.
 Since the $\phi$-theory has a scaling behaviour
(is critical) as seen in the perturbative approach, we can 
write $\log{\cal Z}$ as a power-law in $\Lambda$,
plus an analytical function F($\Lambda$), such that
\begin{equation}\label{Zsca2}
{1 \over V} \; \log{\cal Z}(\Lambda) = {K \over{(2-\alpha)(1-\alpha)}}\;
\Lambda^{2-\alpha} + F(\Lambda) 
\end{equation}
where $ V \sim R^3 $ stands for the volume, $ K$ is a constant
 and alpha is the thermal critical exponent

Calculating the second derivative of $\log{\cal Z}(\Lambda)$ with
respect to $ \Lambda $ (at constant $V$) from eqs.(\ref{Zsca1}) and 
(\ref{Zsca2}) and equating the results, yields
\begin{equation}\label{flucM}
{{\partial^2}\over{\partial\Lambda^2}}\log{\cal Z}(\Lambda) \sim
K V \Lambda^{-\alpha}  \sim R^{{2}\over{\nu}} = 
\int d^3x\; d^3y\;\left[ <\rho({\vec x})\rho({\vec y})>- 
\; <\rho({\vec x})><\rho({\vec y})> \right]
\end{equation}
where we used that $ \Lambda \sim R^{-1/\nu} $ and the scaling
relation $ \alpha = 2 - 3 \nu $ \cite{dg}. 
The r.h.s. of eq.(\ref{flucM}) precisely yields 
the  mass fluctuations  squared $ (\Delta M(R))^2 \equiv  \; <M^2> -<M>^2 $. 
Hence, 
$$
\Delta M(R) \sim R^{d_H} \; .
$$
Therefore, the scaling exponent $\nu$ can be identified with the inverse
Haussdorf (fractal) dimension $d_H$ of the system
$$
d_H = \frac1{\nu} \; .
$$
On one side, the perturbative calculation  yields the mean field value
for $ \nu $ \cite{ll}. That is,
\begin{equation}\label{meanF}
 \nu= \frac12  \quad ,  \quad d_H = 2 \quad {\rm and } \quad q = \frac12 \; .
\end{equation}
On the other side, the renormalization group transformation amounts to replace 
the parameter $ \mu^2 $ in $ S[\phi] $ by the effective one at the scale
in question. This approach
indicates that the long distance critical behaviour is governed by the
(non-perturbative) Ising fixed point  \cite{kgw,dg}. 
Very probably, there are no further fixed points \cite{grexa}. 
The scaling exponents associated to the Ising fixed point are
\begin{equation}\label{Isint}
\nu = 0.631 \quad , \quad d_H = 1.585   \quad {\rm and} \quad
 q = 0.293\; \; .
\end{equation}
From the renormalization group analysis, the two-points density correlation 
function behaves as $ r^{\frac2{\nu} -6}$ or $r^{-2.830}$ for large distances 
($r^{-2}$ for mean field).
This should be compared with observations. Previous attempts to
derive correlation functions from observations were not entirely conclusive, 
because of lack of dynamical range \cite{klein}, but much more extended maps of
the ISM could be available soon to test our theory. In addition, we predict 
an independent exponent for the gravitational potential correlations 
($ \sim r^{-1-\eta} $, where $\eta_{Ising}=0.037$ or $\eta_{mean field}=0$ 
\cite{dg}), which could be checked through 
gravitational lenses observations in front of quasars.

The mean field exponents describe the situation where
non-linear field fluctuations are negligible. When  non-linear
fluctuations are strong, the renormalization 
group exactly accounts for their contributions, giving the
Ising fixed point exponents. 

If we consider the mass of the particles to be the neutral
hydrogen atom, at $T\sim 3K$, and we estimate the fugacity $z$ using the ideal
gas value $z = \rho_0 ({{h^2}\over{2\pi m kT}})^{3/2}$, 
we find the length $\mu^{-1} \sim$ 30 AU (4.5 10$^{14}$ cm), 
and the dimensionless coupling $g^2 \sim 5\; 10^{-53}$, for
a density $\rho_0\sim 10^{10}$ atom/cm$^3$ (\cite{pc}).
This extremely small $g$ supports the perturbative method at these
scales implying the mean field values for the exponents 
(\ref{meanF}). However, the effective coupling constant $g$ grows with the 
scale, according to the renormalization
group flow (towards the Ising fixed point);
$\mu^{-1} $ indicates the order of the smallest distance where the scaling 
regime applies, and corresponds to the observed smallest gravitational scale.
Both  numerical values for the critical exponents (\ref{meanF}) and
(\ref{Isint}) are compatible with the values observed in the present
interstellar medium eq.(\ref{expos}). Further theoretical work in the
$\phi$-theory will determine whether the scaling behaviour is given by the
mean field or by the Ising fixed point.

\bigskip

We have considered for the ISM the simplified view of an isothermal  
self-gravitating gas. This idealized view corresponds exactly
to the outer parts of galaxies, far from any star formation and 
heating sources. There, the molecular cloud ensemble must be in 
isothermal equilibrium with the cosmic background radiation at 
T$\sim$ 3K (e.g. \cite{pcm,pc}). Well inside the galaxy, 
the physics of the ISM is much more 
complex, especially when the violent perturbations due to star 
formation are taken into account. Locally, the ISM around star 
formation regions can effectively lose its fractal structure, 
at least partially (it becomes diffuse and much less fragmented, more or 
less ionized). But radiative cooling is very 
efficient, and shock waves are highly dissipative, so that globally on 
large-scale, the interstellar medium can still be considered as 
isothermal. The bulk of the ISM is only 
slightly perturbed, and we have shown that the scaling laws are 
{\bf stable} under perturbations, so that we believe that
our theory applies to most of the ISM (and in particular to the low 
star-forming Taurus region of figure 1).

Turbulence is probably relevant in the dynamics of the ISM \cite{obser}, 
but this could be a consequence of the fractal structure built
up by gravitational instabilities. It has been recognized for
a long time that the size line-width relation in the ISM is similar
to the Kolmogorov law $ \Delta v \propto  R^{1/3} $ derived for
incompressible subsonic turbulence (this assumes that the energy flow 
per unit mass
($\epsilon \propto \Delta v^3/R $) is constant all over the hierarchy, 
and is finally dissipated on the smallest scales through viscous 
processes). The energy would be powered at large scale by the 
galactic rotation \cite{fle}.
Gravitationally driven compressible turbulence \cite{hen} as well as
gravitational clouds in quasistatic virial equilibrium \cite{chi}
yield mean field exponents  (\ref{meanF}).

The important point demonstrated here is that self-gravity alone
can account for the fractal structure of the ISM, and 
quantitatively predicts its fractal dimension and related 
critical exponents. A new and unexpected connection between the ISM
and  critical phenomena uncovers. 
It is interesting to note that the gravitational gas has been found at 
critical conditions, with correlations at {\bf all scales}, and scale-
independent power-law relations for a {\bf continuous}  range of physical 
parameters (temperature, coupling constant), while the spin models 
with which we have found an analogy, are critical only at a single
value of the temperature.  This feature is
connected with the scale invariant character of the Newtonian force
and its infinite range, i.e. $ r^{-2} $.
 
A further step in the study of the ISM will be to include the
dynamical (time dependent) description within the field theory
approach presented in this paper.
\smallskip

\begin{figure}
  \vspace{0.0 cm}
\caption[]{ Image of the interstellar medium near the Taurus region
at 100$\mu$ wavelength, obtained with the IRAS satellite (the continuum
emission comes from dust heated by the interstellar radiation field, in this
low star-forming region). The linear size of the box is 30 degrees, and the 
pixel is 6 arcmin. A zoom by a factor 3 in scale is made of the central region:
 the overall fragmented structure of the medium remains unchanged. This 
self-similar structure has been confirmed by the compilation of many other 
observations at widely different scales, revealing the fractal structure
of the ISM. }
\end{figure}

\hbox{\epsfxsize 14cm\epsffile{taurus10.cps}}

\hbox{\epsfxsize 14cm\epsffile{taurus30.cps}}

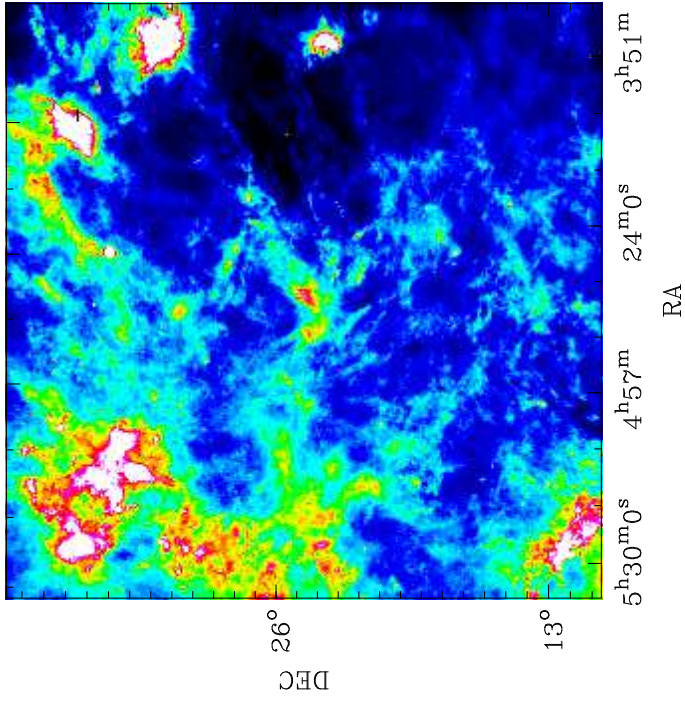
\end{document}